\documentclass[12pt]{article} 
\usepackage{amsmath,amssymb,bbm}
\usepackage{axodraw,subfigure,caption2}
\usepackage[dvips]{graphicx}
\usepackage{cite}

\def\Acal{{\cal A}}
\def\Dcal{{\cal D}}
\def\Kcal{{\cal K}}
\def\Mcal{{\cal M}}
\def\Ocal{{\cal O}}

\def\Ucal{{\cal U}}
\def\Vcal{{\cal V}}
\def\Wcal{{\cal W}}

\def\cc{\text{\sc c}}
\def\ckm{\text{\sc ckm}}
\def\le{\text{\sc l}}

\DeclareMathOperator{\diag}{diag}
\DeclareMathOperator{\trace}{tr}

\def\eabg{\epsilon_{\alpha\beta\gamma}}
\def\abs#1{\left| #1\right|}                    
\def\VEV#1{\left\langle #1\right\rangle}        
\def\braket#1#2#3{\left\langle #1\left| #2\right| #3\right\rangle} 
\def\slash#1{\rlap{\hbox{$\mskip 1 mu /$}}#1}   
\def\kslash{\slash{k}}                    
\def\blb{\Bigl\lbrack}                    
\def\brb{\Bigl\rbrack}                    
\def\Blb{\biggl\lbrack}                   
\def\Brb{\biggl\rbrack}                   

\def\neutrino{k}
\def\charge{\pm}
\def\ina{u_L}
\def\inb{d_L}
\def\inc{s_L}
\def\finalstate#1#2#3{
  \gdef\ina{#1}
  \gdef\inb{#2}
  \gdef\inc{#3}
}

\def\boxud#1#2#3{
  \scalebox{0.95}{
    \begin{picture}(130,85)(20,3)
      \SetWidth{0.8}
      \ArrowLine(20,70)(40,70)  \Text(10,70)[]{$\ina$}
      \ArrowLine(20,30)(40,30)  \Text(10,30)[]{$\inb$}
      \ArrowLine(100,70)(80,70) \Text(110,70)[]{$\nu_{\neutrino\,\le}$}
      \ArrowLine(100,30)(80,30) \Text(110,30)[]{$\inc$} 
      \DashLine(40,70)(80,70)3  \Text(60,80)[]{$\widetilde #1$}
      \DashLine(40,30)(80,30)3  \Text(60,37)[]{$\widetilde #2$}
      \ArrowLine(40,50)(40,70)  \Text(26,50)[]{$\widetilde H_\cc$}
      \ArrowLine(40,50)(40,30)   
      \Line(37,47)(43,53)
      \Line(37,53)(43,47)
      \ArrowLine(80,50)(80,70)  \Text(95,53)[]{$\widetilde{\boson}^\charge$}
      \ArrowLine(80,50)(80,30)
      \Line(77,47)(83,53)
      \Line(77,53)(83,47)
      \Text(0,10)[l]{\small $#3$}
    \end{picture}
    }
  }

\def\cisboxud#1#2#3{
  \scalebox{0.95}{
    \begin{picture}(130,85)(20,3)
      \SetWidth{0.8}
      \ArrowLine(20,70)(40,70)  \Text(10,70)[]{$\ina$}
      \ArrowLine(20,30)(40,30)  \Text(10,30)[]{$\inb$}
      \ArrowLine(100,70)(80,70) \Text(110,70)[]{$\nu_{\neutrino\,\le}$}
      \ArrowLine(100,30)(80,30) \Text(110,30)[]{$\inc$} 
      \DashLine(40,70)(80,70)3  \Text(60,80)[]{$\widetilde #1$}
      \DashLine(40,30)(80,30)3  \Text(60,37)[]{$\widetilde #2$}
      \ArrowLine(40,50)(40,70)  \Text(26,50)[]{$\widetilde{\boson}^\charge$}
      \ArrowLine(40,50)(40,30)   
      \Line(37,47)(43,53)
      \Line(37,53)(43,47)
      \ArrowLine(80,50)(80,70)  \Text(95,53)[]{$\widetilde H_\cc$}
      \ArrowLine(80,50)(80,30)
      \Line(77,47)(83,53)
      \Line(77,53)(83,47)
      \Text(0,10)[l]{\small $#3$}
    \end{picture}
    }
  }

\def\triud#1#2#3{
  \scalebox{0.95}{
    \begin{picture}(130,85)(20,3)
      \SetWidth{0.8}
      \ArrowLine(20,70)(30,50)  \Text(10,70)[]{$\ina$}
      \ArrowLine(20,30)(30,50)  \Text(10,30)[]{$\inb$}
      \ArrowLine(100,70)(80,70) \Text(110,70)[]{$\nu_{\neutrino\,\le}$}
      \ArrowLine(100,30)(80,30) \Text(110,30)[]{$\inc$}
      \DashLine(30,50)(60,50)3  \Text(45,60)[]{$H_\cc$}
      \Line(42,47)(48,53)
      \Line(42,53)(48,47)
      \DashLine(60,50)(80,70)3  \Text(65,35)[]{$\widetilde #2$}
      \DashLine(60,50)(80,30)3  \Text(65,65)[]{$\widetilde #1$}
      \ArrowLine(80,50)(80,70)  \Text(95,53)[]{$\widetilde{\boson}^\charge$}
      \ArrowLine(80,50)(80,30)
      \Line(77,47)(83,53)
      \Line(77,53)(83,47)
      \Text(0,10)[l]{\small $#3$}
    \end{picture}
    }
  }

\def\cistriud#1#2#3{
  \scalebox{0.95}{
    \begin{picture}(130,85)(20,3)
      \SetWidth{0.8}
      \ArrowLine(20,70)(40,70)  \Text(10,70)[]{$\ina$}
      \ArrowLine(20,30)(40,30)  \Text(10,30)[]{$\inb$}
      \ArrowLine(40,50)(40,70)  \Text(26,50)[]{$\widetilde{\boson}^\charge$}
      \ArrowLine(40,50)(40,30)   
      \ArrowLine(100,70)(90,50) \Text(110,70)[]{$\nu_{\neutrino\,\le}$}
      \ArrowLine(100,30)(90,50) \Text(110,30)[]{$\inc$}
      \DashLine(60,50)(90,50)3  \Text(78,60)[]{$H_\cc$}
      \Line(72,47)(78,53)
      \Line(72,53)(78,47)
      \DashLine(40,70)(60,50)3  \Text(58,35)[]{$\widetilde #2$}
      \DashLine(40,30)(60,50)3  \Text(58,65)[]{$\widetilde #1$}
      \Line(37,47)(43,53)
      \Line(37,53)(43,47)  
      \Text(0,10)[l]{\small $#3$}
    \end{picture}
    }
  }

\begin{document}

\title{{
    \normalsize
    \mbox{ }\hfill
    \begin{minipage}{6cm}
      DESY 02-229\\
      hep-ph/0302272\\
      Nucl. Phys. B {\bf 661} (2003) 62-82.
    \end{minipage}}\\
  \vspace{3cm}
  \bf Proton Decay in a Consistent Supersymmetric 
  {\sf SU(5)} GUT Model\\
  \phantom{line}} 

\author{D.~Emmanuel-Costa and S. Wiesenfeldt\\
  \phantom{line}\\
  {\sl Deutsches Elektronen-Synchrotron DESY, 
    22603 Hamburg, Germany}\\
  \phantom{line}}
\date{}

\maketitle

\thispagestyle{empty}

\begin{abstract}
  \noindent
  It is widely believed that minimal supersymmetric {\sf SU(5)} GUTs
  have been excluded by the SuperKamiokande bound for the proton decay
  rate.  In the minimal model, however, the theoretical prediction
  assumes unification of Yukawa couplings, \mbox{$Y_d=Y_e$}, which is
  known to be badly violated.  We analyze the implications of this
  fact for the proton decay rate.  In a consistent {\sf SU(5)} model
  with higher dimensional operators, where {\sf SU(5)} relations among
  Yukawa couplings hold, the proton decay rate can be several orders
  of magnitude smaller than the present experimental bound.

  \medskip
  

\end{abstract}

\newpage


\section{Introduction}

Supersymmetric Grand Unified Theories (SUSY GUTs) \cite{su5} provide a
beautiful framework for theories beyond the standard model (SM) of
particle physics.  They combine several attractive ideas, namely
supersymmetry and unification of matter and interactions.  A crucial
prediction of SUSY GUTs is the instability of the proton
\cite{review}, and the long-lasting search for proton decay has put a
strong constraint on unified theories.

The simplest models are based on the gauge group {\sf SU(5)}.  The SM
particles can be grouped into two multiplets per generation, no
additional matter particles are needed.  Hereby, the down quark and
charged lepton Yukawa couplings are unified.  The GUT scale is set by
the unification of the gauge couplings around $2\times 10^{16}$\,GeV
in the Minimal Supersymmetric Standard Model (MSSM).

{\sf SU(5)} based models have been studied in great detail.  Recently
the simplest version, minimal supersymmetric {\sf SU(5)} \cite{su5},
was claimed to be excluded due to the SuperKamiokande bound on proton
decay \cite{goto99,murayama02}.  The exclusion of the ``prototype" GUT
model is an important result and it is worth analyzing the underlying
assumptions carefully.

One ingredient is the sfermion mixings \cite{bajc02} which are
essentially unknown and which are neglected in
refs.~\cite{goto99,murayama02}.  Taking these mixings into account one
can suppress the proton decay rate below the experimental bound
\cite{bajc02,bajc02b}.  Another important question concerns the
failure of down quark and charged lepton Yukawa couplings to unify.
To our knowledge, all previous analyses assumed exact unification at
the GUT scale, $Y_d=Y_e$, and then used the down quark matrix to study
proton decay.  The decay width, however, is strongly dependent on
flavour mixing and there is no reason not to take, for instance, the
lepton matrix instead.

The failure of Yukawa unification can be accounted for by adding
operators induced by Planck scale effects \cite{bajc02b}.  Since the
GUT scale is only about two orders below the Planck scale, differences
between down quarks and charged leptons can be explained by such
operators.  In addition, they also affect the proton decay operators.

In this paper, we start with minimal supersymmetric {\sf SU(5)} and
discuss the influence of flavour mixing on proton decay.  After that,
we will study the impact of higher dimensional operators on proton
decay.  In particular, we consider two simple models where the decay
rate is well below the experimental limit.

The outline of the paper is as follows: 
After briefly describing the supersymmetric {\sf SU(5)} GUT model 
(Section \ref{su5}) 
and analyzing the dimension five operators (Section \ref{dim5}), 
we discuss the results of the different scenarios in 
Section \ref{results}.
Important and clarifying details are given in the Appendices.


\section{Supersymmetric {\sf SU(5)} GUTs
  \label{su5}}

We start this section by briefly describing the minimal supersymmetric
{\sf SU(5)} GUT model \cite{su5}.  It contains three generations of
chiral matter multiplets,
\begin{align*}
  \mathsf{10}_j &= (Q,u^\cc,e^\cc)_j \ , \\
  \mathsf{5}^*_j &= (d^\cc,L)_j \ ,
\end{align*}
and a vector multiplet $A(\mathsf{24})$ which includes the twelve
gauge bosons of the SM and twelve additional ones, the $X$ and $Y$
bosons.  Because of their electric and colour charges, the latter
mediate proton decay via $d=6$ operators.  At the GUT scale, {\sf
  SU(5)} is broken to ${\sf G_{SM} = SU(3)\times SU(2)\times U(1)_Y}$
by an adjoint Higgs multiplet $\Sigma(\mathsf{24})$.  A pair of
quintets, $H(\mathsf{5})$ and $\overline H(\mathsf{5}^*)$, then breaks
${\sf G_{SM}}$ to ${\sf SU(3)\times U(1)_\text{em}}$ at the
electroweak scale.  The superpotential is given by
\begin{align} 
  \label{minimal-superpotential}
  \begin{split}
    W & = 
    \tfrac{1}{2} m\trace\Sigma^2
    + \tfrac{1}{3} a\trace\Sigma^3
    + \lambda \overline{H}(\mathsf{5}^*) 
    \,(\Sigma + 3 \sigma)\, H(\mathsf{5}) \\
    & \quad 
    + \tfrac{1}{4}\, Y_1^{ij}\,\mathsf{10}_i\;\mathsf{10}_j\;
    H(\mathsf{5})  
    + \sqrt{2}\, Y_2^{ij}\,\mathsf{10}_i\; \mathsf{5}^*_j\;
    \overline{H}(\mathsf{5}^*) 
    \ .
  \end{split}
\end{align}
The adjoint Higgs multiplet,
\begin{align*}
  \Sigma(\mathsf{24}) = \left( \begin{array}{cc}
      \Sigma_8 & \Sigma_{(3,2)} \\
      \Sigma_{(3^*,2)} & \Sigma_3
    \end{array} \right)
  + \frac{1}{2\sqrt{15}} \left( \begin{array}{cc}
      2 & 0 \\ 0 & -3 
    \end{array} \right) \Sigma_{24}\ ,
\end{align*}
acquires the vacuum expectation value (VEV)
\begin{align*}
  \VEV{\Sigma}=\sigma \diag (2,2,2,-3,-3)\ ,
\end{align*}
so that the X and Y bosons become massive,
\begin{equation}
  \label{x-mass}
  M_V \equiv M_X = M_Y = 5 \sqrt{2} g_5 \sigma \ ,
\end{equation}
whereas the SM particles remain massless.  Here $g_5$ is the {\sf
  SU(5)} gauge coupling.  The components $\Sigma_8$ and $\Sigma_3$ of
$\Sigma(\mathsf{24})$ both acquire the mass
\begin{equation*} 
  M_\Sigma\equiv M_8 = M_3 = \tfrac{5}{2} m \ , 
\end{equation*}
while $\Sigma_{(3,2)}$ and $\Sigma_{(3^*,2)}$ form vector multiplets
of mass $M_V$ together with the gauge multiplets.  Finally, the mass
of the singlet component $\Sigma_{24}$ is $\tfrac{1}{2} m$.

The pair of quintets, $H(\mathsf{5})$ and
$\overline{H}(\mathsf{5}^*)$, contains the SM Higgs doublets, $H_f$
and $\overline{H}_f$, which break $\mathsf{G_{SM}}$, and colour
triplets, $H_\cc$ and $\overline{H}_\cc$, respectively.  To have
massless Higgs doublets $H_f$ and $\overline{H}_f$, while their
colour-triplet partners (leptoquarks) are kept super-heavy,
\begin{equation} 
  \label{hc-mass}
  M_{H_\cc} = M_{\overline{H}_\cc} = 5 \lambda \sigma \ ,
\end{equation}
the mass parameters of $H(\mathsf{5})$ and
$\overline{H}(\mathsf{5}^*)$ have to be fine-tuned $\Ocal
(\frac{v}{\sigma}) \sim 10^{-13}$.  This is the so-called
doublet-triplet-splitting problem.  As we will see below, RGE analysis
gives constraints on the masses of the new particles.

Expressed in terms of SM superfields, the Yukawa interactions are
\begin{align}
  \label{SM-superpotential}
  \begin{split}
    W_Y = \;
    & Y_u^{ij}\, Q_i\, u^\cc_j\, H_f 
    + Y_d^{ij}\, Q_i\, d_j^\cc\, \overline{H}_f 
    + Y_e^{ij}\, e_i^\cc\, L_j\, \overline{H}_f \\
    & + \tfrac{1}{2} Y_{qq}^{ij}\, Q_i\, Q_j\, H_\cc 
    + Y_{ql}^{ij}\, Q_i\, L_j\, \overline{H}_\cc 
    + Y_{ue}^{ij}\, u_i^\cc\, e^\cc_j\, H_\cc  
    + Y_{ud}^{ij}\, u_i^\cc\, d_j^\cc\, \overline{H}_\cc 
    \ , 
  \end{split}
\end{align}
where 
\begin{gather}
  \label{up-unification}
  Y_u=Y_{qq}=Y_{ue}=Y_1 \ , \\
  \label{down-unification}
  Y_d=Y_e=Y_{ql}=Y_{ud}=Y_2 \ .
\end{gather}
In particular the Yukawa couplings of down quarks and charged leptons
are unified.  While $m_b=m_\tau$ can be fulfilled at the GUT scale, it
fails for the first and second generation.  This problem can be solved
by adding higher dimensional operators due to physics at the Planck
scale so that \cite{bajc02b}
\begin{equation} 
  \label{higgs-planck}
  W_{\Sigma}= \tfrac{1}{2}m\trace\Sigma^2 
    + \tfrac{1}{3}a\trace\Sigma^3
    + b \frac{(\trace\Sigma^2)^2}{M_\text{Pl}} 
    + c \frac{\trace\Sigma^4}{M_\text{Pl}} \ . 
\end{equation}
Now the masses of $\Sigma_3$ and $\Sigma_8$ are no longer identical,
which will affect the constraints on the leptoquark mass.  Including
possible couplings up to order $1/M_\text{Pl}$, the Yukawa
interactions read
\begin{align}
  \label{yukawa-planck}
  \begin{split}
    W_Y &
    = \frac{1}{4}\,\epsilon_{abcde}\left(
      Y_1^{ij}\, \mathsf{10}_i^{ab}\, \mathsf{10}_j^{cd}\, H^e 
      + f_1^{ij}\, \mathsf{10}_i^{ab}\, \mathsf{10}_j^{cd}\,
      \frac{\Sigma^e_f}{M_\text{Pl}}\,H^f 
      + f_2^{ij}\, \mathsf{10}_i^{ab}\, \mathsf{10}_j^{cf}\, H^d\,
      \frac{\Sigma^e_f}{M_\text{Pl}} 
    \right)\\
    & \quad 
    + \sqrt{2}\left(
      Y_2^{ij}\, \overline{H}_a\, \mathsf{10}_i^{ab}\, 
      \mathsf{5}^*_{jb} 
      + h_1^{ij}\, \overline{H}_a\, \frac{\Sigma^a_b}{M_\text{Pl}}\,
      \mathsf{10}_i^{bc}\, \mathsf{5}^*_{jc} 
      + h_2^{ij}\, \overline{H}_a\, \mathsf{10}_i^{ab}\, 
      \frac{\Sigma_b^c}{M_\text{Pl}}\, \mathsf{5}^*_{jc} 
    \right) \ .
  \end{split}
\end{align}
Then the Yukawa couplings are given by
\begin{align}
  \label{sm-planck}
  \begin{split}
    Y_u & = Y_1 
    + 3\frac{\sigma}{M_\text{Pl}}f_1^S
    + \frac{1}{4}\frac{\sigma}{M_\text{Pl}}\left(3f_2^S+5f_2^A\right)
    \ , \\ 
    Y_d & = Y_2 - 3\frac{\sigma}{M_\text{Pl}}h_1
    + 2\frac{\sigma}{M_\text{Pl}}h_2 
    \ , \\
    Y_e & = Y_2 -3\frac{\sigma}{M_\text{Pl}}h_1
    - 3\frac{\sigma}{M_\text{Pl}}h_2 
    \ . 
  \end{split}
\end{align}
Here $\sigma/M_\text{Pl}\sim\Ocal(10^{-2})$, and $S$ and $A$ denote
the symmetric and antisymmetric parts of the matrices, respectively.
Thus the three Yukawa matrices, which are related to masses and mixing
angles at $M_Z$ by the RGEs, are determined by six matrices.

From eqn.~(\ref{sm-planck}) one reads off, 
\begin{align}
  \label{c-relation}
  Y_d - Y_e = 5\frac{\sigma}{M_\text{Pl}}h_2 \ .
\end{align}
Hence the failure of Yukawa unification is naturally accounted for by
the presence of $h_2$.  Note that we do not need to introduce any
additional field at $M_\text{GUT}$ to obtain this relation; it just
arises from corrections $\Ocal(\sigma/M_\text{Pl})$.  Therefore we
call this model a {\em consistent} supersymmetric {\sf SU(5)} GUT
model.  In the minimal model, $Y_{qq}=Y_{ue}=Y_u$; furthermore, one
usually chooses $Y_{ql}=Y_{ud}=Y_d$.  Note, however, that the choices
$Y_{ql}=Y_{ud}=Y_e$ or $Y_{ql}=Y_d$, $Y_{ud}=Y_e$ would be equally
justified.  As we shall see, this ambiguity strongly affects the
proton decay rate.

\medskip

Finally, in general right-handed neutrinos can be added as singlets in
{\sf SU(5)} models.
With ${\mathsf 1}_j=\nu^\cc_j$,
the Yukawa interactions read
\begin{equation*}
  W_Y^\nu =
  Y_\text{\sc d}^{ij}\,\mathsf{1}_i\; \mathsf{5}^*_j\; H(\mathsf{5})
  + M^{ij}\,\mathsf{1}_i\; \mathsf{1}_j \ ,
\end{equation*}
where $M$ is the Majorana mass matrix with eigenvalues
$\Ocal(M_\text{GUT})$. 


\section{Analysis of dimension five operators 
  \label{dim5}}

The evolution of the proton decay rate based on dimension five
operators involves a number of parameters and assumptions which have
changed in analyses during the past years. In this section we
therefore list the main ingredients of our quantitative
analysis. Technical details are given the Appendices.

\medskip

\begin{figure}
  \centering
  \subfigure[]
  { 
    \label{decay:a}
    \begin{picture}(200,60)(20,5)
      \ArrowLine(30,50)(80,30)   \Text(22,53)[]{$u_\le$}
      \ArrowLine(30,10)(80,30)   \Text(22,12)[t]{$d_\le$}
      \Vertex(80,30)2            \Text(80,21)[t]{$LLLL$}
      \DashLine(80,30)(130,50)5  \Text(115,60)[t]{$\widetilde l$}
      \DashLine(80,30)(130,10)5  \Text(115,12)[t]{$\widetilde q$}
      \ArrowLine(160,50)(130,50) \Text(172,52)[t]{$\nu_\le$}
      \Vertex(130,50)1           
      \Line(130,10)(130,50)
      \Photon(130,10)(130,50){2}{6}
      \Text(142,30)[]{$\widetilde{w}^\charge$} 
      \Vertex(130,10)1           
      \ArrowLine(160,10)(130,10) \Text(169,12)[t]{$s_\le$}
    \end{picture}
    }
  \subfigure[]
  { 
    \label{decay:b}
    \begin{picture}(180,60)(20,5)
      \ArrowLine(30,50)(80,30)   \Text(22,53)[]{$u^\cc$}
      \ArrowLine(30,10)(80,30)   \Text(22,12)[t]{$d^\cc$}
      \Vertex(80,30)2            \Text(80,21)[t]{$RRRR$}
      \DashLine(80,30)(130,50)5  \Text(110,48)[]{$\widetilde e^\cc$}
      \DashLine(80,30)(130,10)5  \Text(110,27)[]{$\widetilde u^\cc$}
      \ArrowLine(160,50)(130,50) \Text(172,52)[t]{$\nu_\le$}
      \Vertex(130,50)1           \Text(130,57)[]{$y$}
      \ArrowLine(130,30)(130,50)
      \ArrowLine(130,30)(130,10) \Vertex(130,30)1
      \Text(142,30)[]{$\widetilde{h}^\charge$} 
      \Vertex(130,10)1           \Text(128,1)[]{$y^\prime$}
      \ArrowLine(160,10)(130,10) \Text(169,12)[t]{$s_\le$}
    \end{picture}
    }
  \caption{Proton Decay via dimension five operators:
    They result from exchange of the leptoquarks followed by gaugino
    or higgsino dressing.}
  \label{decay}
\end{figure}
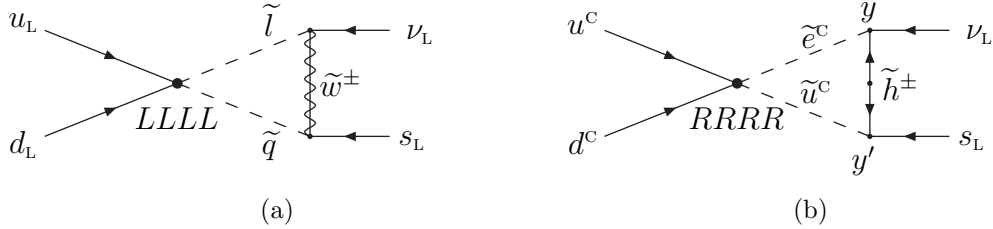  

Integrating out the leptoquarks in eqn.~(\ref{SM-superpotential}),
two dimension five operators remain which lead to proton decay
(fig.~\ref{decay}) \cite{dim5op},
\begin{equation}
  \label{operator}
  W_5 =
  \frac{1}{M_{H_\cc}} \left[
  \, \frac{1}{2} \, Y_{qq}^{ij}\, Y_{ql}^{km}\, \left( 
    Q_i\,Q_j\right)\left( 
    Q_k\,L_m\right)
  + Y_{ue}^{ij}\, Y_{ud}^{km}\, \left(
    u_i^\cc\,e_j^\cc\right)\left(
    u_k^\cc\,d_m^\cc\right)\,\right] \ , 
\end{equation}
called the $LLLL$ and $RRRR$ operator, respectively.  The scalars are
transformed to their fermionic partners by exchange of a gauge or
Higgs fermion.  Neglecting external momenta, the triangle diagram
factor reads, up to a coefficient $\Kcal$ depending on the exchange
particle,
\begin{align} 
  \label{exchange}
  \int \frac{d^4 k}{i (2 \pi)^4}\, \frac{1}{m_1^2 - k^2}\,
  \frac{1}{m_2^2 - k^2}\, \frac{1}{M - \kslash} =
  \frac{1}{(4\pi)^2}  f(M; m_1, m_2) \ ,
  \intertext{with}
  f(M; m_1, m_2) = \frac{M}{m_1^2 - m_2^2} 
  \left( \frac{m_1^2}{m_1^2 - M^2}\, \ln \frac{m_1^2}{M^2} -
  \frac{m_2^2}{m_2^2-M^2}\, \ln \frac{m_2^2}{M^2} \right)\ ,
  \label{triangle}
\end{align}
where $M$ and $m_j$ denote the gaugino and sfermion masses,
respectively. 

As a result of Bose statistics for superfields, the total
anti-symmetry in the colour index requires that these operators are
flavour non-diagonal \cite{dimopoulos82}.  The dominant decay mode is
therefore $p\to K \bar\nu$.  Since the dressing with gluinos and
neutralinos is flavour diagonal, the chargino exchange diagrams are
dominant \cite{nath85,hisano93}.  The wino exchange is related to the
$LLLL$ operator and the charged higgsino exchange to the $RRRR$
operator, so that the coefficients of the triangle diagram factor are 
\begin{equation}
  \label{kfactor}
  \Kcal_{LLLL}=2g^2 \ , \quad
  \Kcal_{RRRR}=y\, y^\prime \ .
\end{equation}
Here $y$ and $y^\prime$ denote the corresponding Yukawa couplings (cf.
fig.~\ref{decay:b}) and g is the gauge coupling.

\medskip

The leading process $p\to K^+ \bar{\nu}$ is used in the analyses of 
Goto and Nihei \cite{goto99} and Murayama and Pierce \cite{murayama02}
to exclude the minimal supersymmetric {\sf SU(5)} model.

\subsubsection*{Calculation of the leading process}

The Wilson coefficients $C_{5L} = Y_{qq} Y_{ql}$ and $C_{5R} = Y_{ue}
Y_{ud}$ are evaluated at the GUT scale.  Then they have to be evolved
down to the scale $M_\text{SUSY}$, leading to a short-distance
renormalization factor $A_s$.  The sparticles are integrated out, as
described above, and the operators give rise to the effective
four-fermion operators of dimension 6.  Now the renormalization group
procedure goes on to the scale of the proton mass $m_p\sim 1\,$GeV
leading to a second, long-distance renormalization factor $A_l$.  The
factors are discussed in Appendix \ref{factor}.

At 1\,GeV, the link to the hadronic level is made using the chiral
Lagrangian method \cite{claudson,chadha}.  In ref.~\cite{goto99}, the
Amplitude for $p \to K^+\bar{\nu}$ is given as
\begin{align}
  \label{amplitude}
  \begin{split}
    \Acal (K^+ \bar\nu) 
    & =
    \Blb \beta\, C^{usd\nu}_{LL}\! + \alpha\, C^{usd\nu}_{RL} \Brb 
    \frac{2 m_p}{3 m_B} D\! +\! 
    \Blb \beta\, C^{uds\nu}_{LL}\! + \alpha\, C^{uds\nu}_{RL} \Brb 
    \left( 1\!+\!\frac{m_p}{3 m_B}(3F\!+\!D) \right) \\
    & \quad + \alpha\, C^{dsu\nu}_{RL} 
    \left( 1\!-\!\frac{m_p}{3 m_B}(3F\!-\!D) \right) \ ,
  \end{split}
\end{align}
where
\begin{itemize}
\item $\alpha$ and $\beta$ are the hadron matrix elements
  \cite{brodsky83} 
  \begin{align}
    \begin{split}
      \alpha\, u_L({\bf k}) = \;
      & \eabg \braket{0}{(d_R^\alpha\, u_R^\beta)\, u_L^\gamma}
      {\,p\,({\bf k})\,} \ , \\ 
      \beta\, u_L({\bf k}) = \;
      & \eabg \braket{0}{(d_L^\alpha\, u_L^\beta)\, u_L^\gamma}
      {\,p\,({\bf k})\,} \ ,
    \end{split}  
  \end{align}
  from which all other elements can be calculated.
  In our analysis we need \cite{jlqcd}
  \begin{align*}
    \braket{K^+}{(u s_R) d_L}{p} &=
    \frac{\alpha}{f_\pi}\,\frac{2\, m_p}{3\, m_B}\, D \ , \\
    \braket{K^+}{(u d_R) s_L}{p} &= \frac{\alpha}{f_\pi}
    \left[ 1+ \left(F+\frac{1}{3}D\right) \frac{m_p}{m_B} \right] \ , 
  \end{align*}
  and $\alpha\leftrightarrow\beta$ for $R\leftrightarrow L$\,.
  $u_L({\bf k})$ denotes 
  the left-handed component of the proton wave function,
  $f_\pi=131\,$MeV the pion decay constant.  It is known that
  $\abs\alpha \simeq \abs\beta$ \cite{brodsky83}, and different
  calculations give $0.003\,\text{GeV}^3\leq\beta\leq
  0.03\,\text{GeV}^3$.  The latest evaluation was done by the JLQCD
  collaboration at 2.3\,GeV obtaining $\alpha=-0.015(1)\,\text{GeV}^3$
  and $\beta=0.014(1)\,\text{GeV}^3$ \cite{jlqcd}.  The systematic
  uncertainties are large, and since we want to study whether the
  experimental limit excludes minimal supersymmetric {\sf SU(5)} or
  not, we use the smallest value $\beta=0.003\,\text{GeV}^3$;
\item $m_B=1150\,$MeV is an average baryon mass according to
  contributions from diagrams with virtual $\Sigma$ and $\Lambda$
  \cite{claudson};
\item $D=0.81$ and $F=0.44$ are the symmetric and antisymmetric
  {\sf SU(3)} reduced matrix elements for the axial-vector current
  \cite{shrock78};
\item the coefficients $C_{LL}$ and $C_{RL}$ are related to the $LLLL$
  and $RRRR$ operators as discussed in Appendix \ref{diagrams} and
  given by eqn.~(\ref{coefficient}).  Moreover, they include the
  renormalization factors $A_s$ and $A_l$ as well as the coefficient
  ${\mathcal{K}}$ (\ref{kfactor}), the triangle diagram factor
  (\ref{triangle}) and the suppressing mass of the leptoquarks,
  ${M_{H_\cc}}$.
\end{itemize}
The first line of eqn.~(\ref{amplitude}) is related to chargino
exchange as shown in fig.~\ref{chargino-exchange}.  This formula is
given in ref.~\cite{lucas97}.  The authors of ref.~\cite{goto99} add
the third term due to neutralino exchange of
fig.~\ref{neutralino-exchange:b}.  Here we also include the
corresponding diagrams of fig.~\ref{neutralino-exchange:a}.

Finally, the decay width is given by \cite{chadha}
\begin{align}
  \label{totalwidth}
  \Gamma(p\to K^+ \bar\nu) =  
  \dfrac{(m_p^2-m_K^2)^2}{32\pi m_p^3 f_\pi^2}\,
  \sum_{i=e,\mu,\tau}
  \abs{\mathcal{{A}}(K^+ \bar\nu_i)}^2 \ .
\end{align}

\subsubsection*{Comparing the $LLLL$ and $RRRR$ contribution}

The $RRRR$ contribution has been ignored for a long time.  However, as
pointed out by Lucas and Raby \cite{lucas97}, this operator gives a
significant contribution in SUSY {\sf SO(10)} models.  The reason is
that the Wilson coefficients and hence the $LLLL$ contribution are
proportional to $\frac{1}{\sin 2\beta}=\frac{1}{2} (\tan\beta +
\frac{1}{\tan\beta})$, whereas the $RRRR$ contribution is proportional
to $(\tan\beta + \frac{1}{\tan\beta})^2$. Therefore the latter is
dominant for large $\tan\beta$, which is naturally the case for {\sf
  SO(10)} models.  Here $\tan\beta$ defines the ratio of the vacuum
expectation values of the Higgs doublets $H_f$ and $\overline H_f$.
Since the $RRRR$ contribution is proportional to the Yukawa couplings,
it is dominated by the third generation.  As long as the top mass was
believed to be less than 100\,GeV, it could be neglected in the
analysis.  Then the decay width is given by the $LLLL$ contribution
and can be suppressed sufficiently by adjusting the phase matrix given
in eqn.~(\ref{phase}).

In ref.~\cite{goto99}, the $RRRR$ contribution was studied in the
minimal {\sf SU(5)} model.  It was found that the total width is even
affected for low $\tan\beta$ because the phase dependence of $p\to K^+
\bar\nu_\mu$ and $p\to K^+ \bar\nu_\tau$ now differs, so both channels
cannot be reduced simultaneously.

\subsubsection*{Supersymmetric particle spectrum}

Looking at the dressing diagram 
we notice that by taking the sfermions to be degenerate at a TeV,
the triangle diagram factor (\ref{triangle}) is given by
\begin{equation} 
  \label{triangle2}
  f(M;m) = 
  \dfrac{M}{(M^2-m^2)^2} \left(m^2-M^2-M^2\, \ln\frac{m^2}{M^2}\right)
  \; \xrightarrow{\; M \ll m\;} \;
  \frac{M}{m^2} \ .
\end{equation}
Therefore the sfermions are usually assumed to have masses of 1\,TeV.
An exception is often made for top squarks.  Since the off-diagonal
entries of the mass matrix are proportional to $m_t$, the mixing is
expected to be large, with at least one eigenvalue much below 1\,TeV.
In analyses, one typically uses 400\,GeV, 800\,GeV, or 1\,TeV for
$m_{\tilde t}$.  For the other sfermions, the mixings are neglected.
The proton decay rate is further suppressed by light gauginos and
higgsinos. Note that the experimental limit for charginos is
$m_{\widetilde{\chi}^\pm}>67.7\,$GeV \cite{pdg}.

Since proton decay is dangerously large, also the decoupling scenario
\cite{decoupl} has been studied, where the scalars of the first and
second generation can be as heavy as 10\,TeV \cite{murayama02}.  Here,
proton decay is clearly dominated by the third generation.

As already mentioned above, one can constrain the leptoquark mass
$M_{H_\cc}$ by examining the RGEs for the gauge couplings; the details
are given in Appendix \ref{rge}.  This analysis has been done in the
minimal model for a long time already, first at one-loop level then at
two-loop accuracy because of the large top Yukawa coupling.  The most
recent calculation leads to the constraint \cite{murayama02},
\begin{equation} 
  3.5\times 10^{14} \,\text{GeV} \leq 
  M_{H_\cc} \leq 3.6\times 10^{15}\,\text{GeV} 
  \quad \text{(90\% C.L.)} \ ,
  \label{newconstraint}
\end{equation}
with $M_{H_\cc}$ well below the GUT scale.

This constraint depends on the Higgs representations.  Other Higgs
representations can be chosen as well which yield a higher leptoquark
mass (cf.~\cite{altarelli}).  Moreover, we already pointed out that
$M_3 = M_8$ no longer holds in the consistent model.  Then $M_{H_\cc}$
changes by a factor of $(M_3/M_8)^{5/2}$ and we easily estimate that
$M_3 = 2\, M_8$ is enough to raise the limit to $M_\text{GUT}$.  In
our calculation, we therefore choose $M_{H_\cc}=2\times 10^{16}$\,GeV.

Finally,
one can also define a quantity $M_\text{GUT}$ 
for which one gets \cite{murayama02}, 
\begin{equation} 
  1.7\times 10^{16} \,\text{GeV} \leq 
  M_\text{GUT} \equiv (M_V^2 M_\Sigma)^{1/3} \leq 
  2.0\times 10^{16}\,\text{GeV} 
  \quad \text{(90\% C.L.)} \ ,
  \label{gutconstraint}
\end{equation}
in good agreement with the region of the gauge coupling
unification.


\section{Minimal versus consistent models
  \label{results}}

In this section we want to discuss the decay rate both in the minimal
and in the consistent {\sf SU(5)} model.  The diagrams are given in
Appendix \ref{diagrams}.
Finally, we calculate the decay via dimension six operators.


\subsection{Minimal model}

As already discussed in Section\,\ref{su5}, we can choose $Y_d$ or
$Y_e$ for $Y_{ql}$ and $Y_{ud}$ to calculate the proton decay
amplitude.  Since the Yukawa couplings of down quarks and charged
leptons do not unify, this ambiguity cannot be resolved in minimal
{\sf SU(5)}.  Despite this fact, however, in all previous analyses the
relations $Y_{ql}=Y_{ud}=Y_d$ have been used.

As discussed in Appendix \ref{basis}, two physical bases are used to
calculate the decay amplitudes, with either a diagonal up quark matrix
\cite{hisano93} or a diagonal down quark matrix \cite{goto99}.
Assuming
\begin{align}
  Y_{qq}=Y_{ue}=Y_u \ , \qquad
  Y_{ql}=Y_{ud}=Y_d \ ,
\end{align}
the Wilson coefficients at the GUT scale can be written as 
\begin{align}
  \label{wilson1}
  \begin{split}
    C_{5L}^u = Y_{qq}^u Y_{ql}^u
    & = (\Dcal_u\, P) (V_\ckm\, \Dcal_d)\ , \\
    C_{5R}^u = Y_{ue}^u Y_{ud}^u
    & = (\Dcal_u\, V_\ckm^*) (P^*\, V_\ckm \Dcal_d)
  \end{split}
\end{align}
in the former and
\begin{align}
  \label{wilson2}
  \begin{split}
    C_{5L}^d = Y_{qq}^d Y_{ql}^d
    & = (V_\ckm^T P\, \Dcal_u V_\ckm) (\Dcal_d)\ , \\
    C_{5R}^d = Y_{ue}^d Y_{ud}^d
    & = (V_\ckm^T \Dcal_u) (P^* V_\ckm^* \Dcal_d)
  \end{split}
\end{align}
in the latter case.  Here $\Dcal_u$ and $\Dcal_d$ are the diagonalized
Yukawa coupling matrices evaluated from $Y_u$ and $Y_d$, respectively,
$V_\ckm$ is the CKM matrix and $P$ is the additional phase matrix
as given in eqn.~(\ref{phase}).
  
\begin{figure}[t]
  \centering
  \includegraphics[width=.7\linewidth]{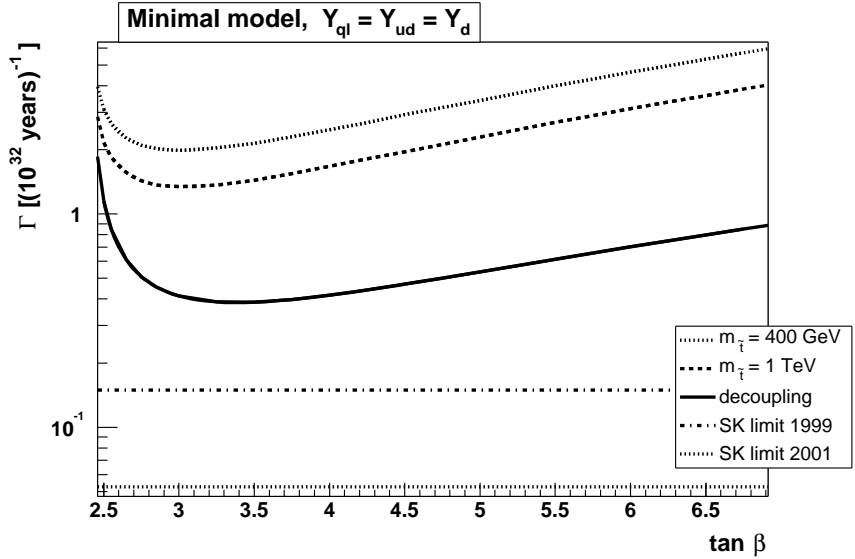}
  \caption{Decay rate $\Gamma(p\to K^+ \bar\nu)$ as function of 
    $\tan\beta$ in the minimal model with 
    \mbox{$Y_{ql}=Y_{ud}=Y_d$}.
    The experimental limits are given by SuperKamiokande experiment
    \cite{hayato99,ganezer01}.}  
  \label{minimal-down}
\end{figure}

We choose the parameters in eqn.~(\ref{totalwidth}) as described in
Section~\ref{dim5} and vary $\tan\beta$.  Since the decay width is
proportional to $\tan\beta$, low values are preferred to obtain a
small decay rate.  On the one hand, the top Yukawa coupling becomes
non-perturbative for low $\tan\beta$ since
$h_t\simeq\frac{1}{\sin\beta}$.  Hence, we start at $\tan\beta\simeq
2.5$.  Fig.~\ref{minimal-down} shows the results of the following
three cases: (i) all sfermions have masses of 1\,TeV; (ii) $m_{\tilde
  t}$ is changed to 400\,GeV; (iii) decoupling scenario, where the
scalars of the first and second generation have masses of 10\,TeV.
The dash-dotted line represents the experimental limit $\tau =
6.7\times 10^{32}$\,years as given by the SuperKamiokande experiment
\cite{hayato99,pdg}, the dotted line is the new limit $\tau =
1.9\times 10^{33}$\,years \cite{ganezer01}.  As in
refs.~\cite{goto99,murayama02}, the amplitude is always above the
experimental limit.

\medskip

\begin{figure}[t]
  \centering
  \includegraphics[width=.7\linewidth]{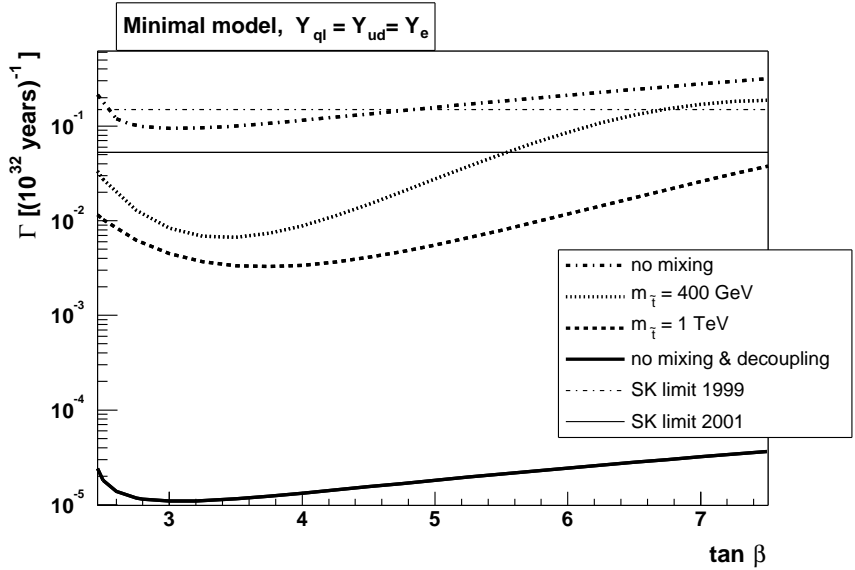}
  \caption{Decay rate $\Gamma(p\to K^+ \bar\nu)$ as a function of 
    $\tan\beta$ with $Y_{ql}=Y_{ud}=Y_e$.
    The mixing matrix $\Mcal$ is taken arbitrary or
    $\Mcal=\mathbbm{1}$.} 
  \label{minimal-electron}
\end{figure}

Next, we study the case 
\begin{align}
  Y_{qq}=Y_{ue}=Y_u \ , \qquad 
  Y_{ql}=Y_{ud}=Y_e \ ,
\end{align}
in order to illustrate the strong dependence of the decay rate on
flavour mixing and therefore on Yukawa unification.  The Wilson
coefficients now read
\begin{align}
  \label{wilson-e1}
  \begin{split}
    C_{5L}^u & = (\Dcal_u\, P) (\Mcal\, \Dcal_e) \ , \\
    C_{5R}^u & = (\Dcal_u\, \Mcal^*) (P^*\, \Mcal \Dcal_e)
  \end{split}
  \intertext{and}
  \begin{split}
  \label{wilson-e2}
    C_{5L}^d & = (\Mcal^T P\, \Dcal_u \Mcal) (\Dcal_e) \ , \\
    C_{5R}^d & = (P^* \Mcal^* \Dcal_e) (\Mcal^T \Dcal_u) \ ,
  \end{split}
\end{align}
where $\Mcal=U_u^\dagger\,U_e$ replaces the CKM matrix $V_\ckm$.  Note
that the mixing matrix in $Y_u$ or $Y_d$ (cf.
eqs.~(\ref{hisano-basis}) and (\ref{goto-basis})) is still given by
$V_\ckm$.  Since $Y_d\not= Y_e$, the masses and mixing of quarks and
leptons are different and $\Mcal$ is undetermined.

We first ignore mixing, i.e. $\Mcal=\mathbbm{1}$, and calculate the
decay rate; the results are shown in fig.~\ref{minimal-electron}.
Without mixing, only scalars of the first and second generation take
part so that the decay rate can be reduced significantly in the
decoupling scenario where the triangle diagram factor (\ref{triangle})
changes by almost two orders of magnitude.

Now we take $\Mcal$ totally arbitrarily and minimize the decay rate.
As can be seen in fig.~\ref{minimal-electron}, it is possible to push
the amplitude below the experimental limit even for smaller sfermion
masses.  In the case $m_{\tilde t}=400$\,GeV, this is only possible
for small values of $\tan\beta$.  The fact that a sufficiently low
decay rate can be found illustrates the dependence on flavour mixing
and therefore the uncertainty due to the failure of Yukawa
unification.


\subsection{Consistent model}

In this case the coefficients of the operators can be derived from the
superpotential (\ref{yukawa-planck}),
\begin{align}
  \begin{split}
    \label{up-planck}
    Y_{qq} & = Y_1
    - 2\frac{\sigma}{M_\text{Pl}}f_1^S 
    - \frac{1}{2}\frac{\sigma}{M_\text{Pl}}f_2^S \ ,  \\ 
    Y_{ue} & = Y_1
    - 2\frac{\sigma}{M_\text{Pl}}f_1^S
    - \frac{1}{2}\frac{\sigma}{M_\text{Pl}}\left(f_2^S+5f_2^A\right)
  \end{split} 
  \intertext{and}
  \begin{split}
    \label{down-planck}
    Y_{ql} & = Y_2 + 2\frac{\sigma}{M_\text{Pl}}h_1
    - 3\frac{\sigma}{M_\text{Pl}}h_2 \ , \\
    Y_{ud} & = Y_2 + 2\frac{\sigma}{M_\text{Pl}}h_1
    + 2\frac{\sigma}{M_\text{Pl}}h_2 \ .
  \end{split}
\end{align}
Note that $Y_{ql}-Y_{ud}=Y_e-Y_d$, which means that $Y_{ql}$ and
$Y_{ud}$ cannot be zero at the same time.

It is instructive to express these Yukawa matrices in terms of the
quark and charged lepton Yukawa couplings and the additional matrices
$f$ and $h$ (cf. relations eqs.~(\ref{consistent-relations})),
\begin{align}
  \begin{split}
    \label{wilson-planck}
    Y_{qq} = Y_{qq}^S = Y_{ue}^S &= Y_u^S 
    - 5\frac{\sigma}{M_\text{Pl}}\left(
      f_1^S+\frac{1}{4} f_2^S\right)\ , \\
    Y_{ue}^A & = Y_u^A 
    - \frac{5}{2}\,\frac{\sigma}{M_\text{Pl}}f_2^A \ , \\
    Y_{ql} & = Y_e + 5\frac{\sigma}{M_\text{Pl}}h_1 \ , \\
    Y_{ud} & = Y_d + 5\frac{\sigma}{M_\text{Pl}}h_1 \ .
  \end{split}
\end{align}
If one allows the (3,3)-component of $f_1$ and $f_2$ to be 
$\Ocal(\frac{M_\text{Pl}}{\sigma})\gg 1$,
proton decay via dimension five operators can be avoided,
for instance, by satisfying
\begin{align}
  f_1^S+\frac{1}{4} f_2^S =
  \frac{M_\text{Pl}}{5\,\sigma}\, Y_u^S \ , \qquad
  f_2^A =
  \frac{2}{5}\,\frac{M_\text{Pl}}{\sigma}\, Y_u^A
\end{align}
so that both $C_{5L}=Y_{qq} Y_{ql}$ and $C_{5R}=Y_{ue} Y_{ud}$
vanish. 

Even if we restrict ourselves to `natural matrices', i.e. couplings
$\Ocal(1)$, we can considerably reduce the decay amplitudes by a
suitable choice of matrices.  In the following, we will illustrate
this with two simple examples where either the $RRRR$ or the $LLLL$
contribution vanishes at the GUT scale.

The first model is given by 
\begin{align}
  \label{consistent-model}
  \begin{split}
    Y_{qq} = Y_{ue} & = \diag(0,0,y_t) \ ,\\
    Y_{ud} & = \diag(0,y_s-y_\mu,y_b-y_\tau) \ ,\\
    Y_{ql} & = \diag(y_e-y_d,0,0) \ ,
  \end{split}
\end{align}
where $y_j$ are the Yukawa couplings of the fermions at
$M_\text{GUT}$.  In this model the $RRRR$ contribution vanishes
completely because $C_{5R}^{ijkm}=Y_{ue}^{ij}\ Y_{ud}^{km}$ is zero
whenever a particle of the first generation takes part.  But according
to figs.~\ref{chargino-exchange:d} and \ref{neutralino-exchange:b}, at
least one particle of the first generation is needed.  Furthermore,
only the decay channel $p\to K\bar\nu_{e}$ remains.  As requested, all
matrix elements are $\Ocal(1)$ or smaller.

\begin{figure}[t]
  \centering
  \includegraphics[width=.7\linewidth]{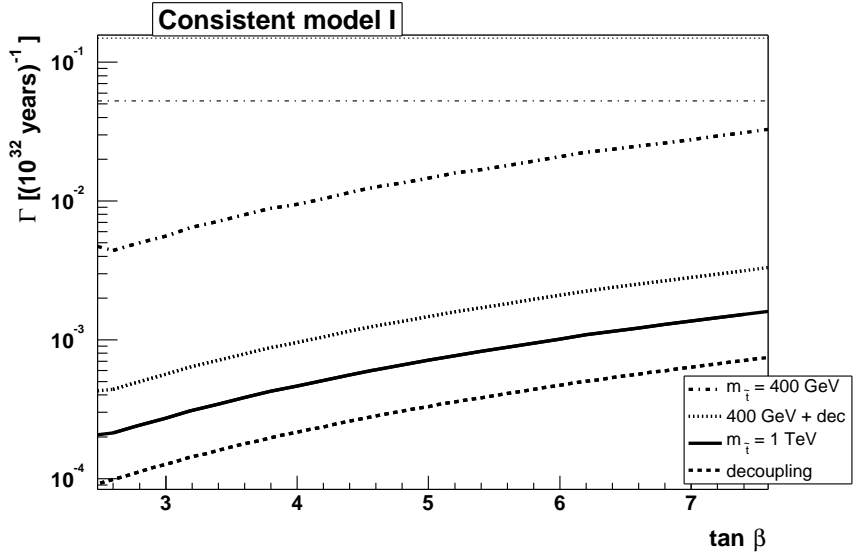}
  \caption{Decay rate $\Gamma(p\to K^+ \bar\nu)$ as function of 
    $\tan\beta$ in the consistent model.
    }
  \label{consistent}
\end{figure}

After RGE evolution by means of eqs.~(\ref{c5Lren}) and
(\ref{c5Rren}), the simple structure of Wilson coefficients changes
slightly, but the $RRRR$ contribution and the decay channel $p\to
K\bar\nu_{\mu}$ are still negligible whereas $p\to K\bar\nu_{\tau}$
becomes dominant.  Fig.~\ref{consistent} shows the results for
different sfermion masses.  The decay amplitude is always well below
the experimental limit, in the case $m_{\tilde t}=1$\,TeV even more
than two orders of magnitude.

\begin{figure}[t]
  \centering
  \includegraphics[width=.7\linewidth]{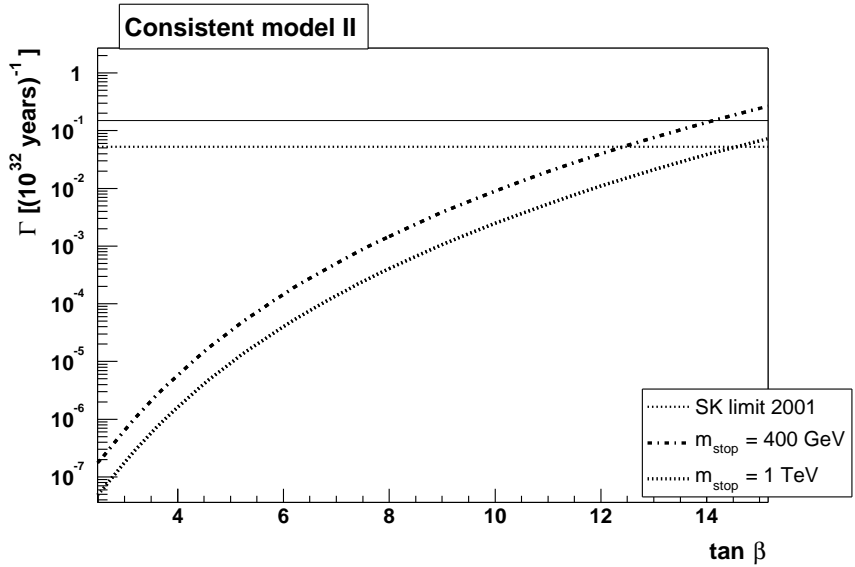}
  \caption{Decay rate $\Gamma(p\to K^+ \bar\nu)$ in the second model.} 
  \label{consistent2}
\end{figure}

\medskip

Now we turn to the second model
where the $LLLL$ contribution vanishes at $M_\text{GUT}$,
\begin{align}
  \begin{split}
    Y_{qq} = Y_{ue} &= \diag(0,0,y_t)\ , \\
    Y_{ud} &= \diag(y_d-y_e,y_s-y_\mu,y_b)\ , \\
    Y_{ql} &= \diag(0,0,y_\tau)\ .
  \end{split}
\end{align}
Now $C_{5L}^{ijkm}=Y_{qq}^{ij}\ Y_{ql}^{km}$ is only different from
zero for $i=j=k=m=3$ and the decay has to be non-diagonal.  Only the
$RRRR$ contribution with a low absolute value remains.  After
renormalization, the $RRRR$ contribution is still dominated by third
generation scalars so that decoupling of the first and second
generation does not change the result.  The $LLLL$ operator
contributes only via $p\to K\bar\nu_{\tau}$.

As shown in fig.~\ref{consistent2}, the proton decay rate is even
smaller in this model.  Furthermore, due to the smaller
(3,3)-component of $h_1$ compared to the first model, it can easily be
used for higher values of $\tan\beta$.

\subsection{Proton decay via dimension six operators}

For completeness we include proton decay via X and Y bosons
\cite{buras78,ellis79}. 
The dominant decay mode is $p\to e^+ \pi^0$.
The decay width is given by
\begin{align}
  \Gamma(p\to e^+ \pi^0) &=  
    \frac{m_p}{64\pi f_\pi^2}\, \alpha^2 
    (1+D+F)^2\, 
    \Bigl( \frac{g_u^2}{M_V^2}A \Bigl)^2\,
    \blb\, 1+ \big(\, 1+\left| V_{ud}\right|^2 \big)^2 \brb
    \;.
\end{align}
The enhancement factor $A$ contains both a short-distance
contribution $A^\text{SD}$ between the SUSY-breaking and GUT scales
and a long-distance contribution $A^\text{LD}\equiv A_l$ between
1\,GeV and the SUSY-breaking scale.  $A^\text{SD}$ splits into three
parts according to the three gauge couplings:
\begin{align}
  A^\text{SD} 
  = \Blb \frac{\alpha_1(M_Z)}{\alpha_5}\Brb^\frac{23}{30\,b_1} 
  \Blb \frac{\alpha_2(M_Z)}{\alpha_5}\Brb^\frac{3}{2\,b_2} 
  \Blb \frac{\alpha_3(M_Z)}{\alpha_5}\Brb^\frac{4}{3\,b_3}
  = 2.37 \ ,
\end{align}
where the first part is an approximate calculation \cite{ibanez84}.

With the Super-Kamiokande limit $\tau(p\to e^+ \pi^0) = 5.3\times
10^{33}$\,years \cite{suzuki} and using $\alpha=0.015\,\text{GeV}^3$
of ref.~\cite{jlqcd}, the mass of the heavy gauge bosons has to
satisfy the lower bound
\begin{align}
  M_V \ge 6.8\times 10^{15}\,\text{GeV}\ ,
\end{align}
roughly half of $M_\text{GUT}$ (\ref{gutconstraint}).  Since
$\tau\propto M_V^4$, the proton decay rate for $M_V=M_\text{GUT}$ is
far below the detection limit which can be reached within the next
years.


\section{Conclusion
  \label{outlook}}

We have recalculated the proton decay rate in supersymmetric {\sf
  SU(5)} GUTs.  In particular, we have emphasized the strong
dependence of the decay amplitude for flavour mixing.

Minimal {\sf SU(5)} GUT is inconsistent since the predicted Yukawa
unification, $Y_d=Y_e$, is badly violated.  A consistent
supersymmetric {\sf SU(5)} model requires additional interactions
which account for the difference of down quark and charged lepton
masses. Such interactions are conveniently parameterized by higher
dimensional operators.

We have shown that such operators can reduce the proton decay rate by
several orders of magnitude and make it consistent with the
experimental upper bound. We are not aware of a mechanism which would
naturally lead to the required relations among Yukawa couplings. But,
on the other hand, proton decay also does not rule out consistent
supersymmetric {\sf SU(5)} models.

\subsubsection*{Acknowledgements}

We are very grateful to Wilfried Buchm\"uller for his guidance in
this project.
The work of D.E.C. was supported by
Funda\c{c}\~{a}o para a Ci\^{e}ncia e a Tecnologia
under the grant SFRH/BPD/1598/2000.

\newpage


\begin{appendix}
  \numberwithin{equation}{section}

  \section{The {\sf SU(5)} Yukawa sector and specific bases 
    \label{basis}}

  \subsubsection*{Minimal model}

  In the minimal theory,
  the {\sf SU(5)} Yukawa sector of the superpotential reads 
  \begin{align*} 
    W_Y 
    = \tfrac{1}{4}\, Y_1^{ij}\,\mathsf{10}_i\; \mathsf{10}_j\;
    H(\mathsf{5}) 
    + \sqrt{2}\, Y_2^{ij}\,\mathsf{10}_i\; \mathsf{5}^*_j\;
    \overline{H}(\mathsf{5}^*)\ . 
  \end{align*}
  From the superpotential one can immediately conclude that $Y_1$ is a
  symmetric complex matrix.  With $Y_u=Y_1$ and $Y_d=Y_2$, the Yukawa
  matrices have the form
  \begin{align*}
    Y_u &= U_u\,\Dcal_u\,P\,U_u^\top\ ,\\
    Y_d &= U_d\,\Dcal_d\,U_{d\,R}^\dagger\ .
  \end{align*}
  Here, $P$ is an additional phase matrix with $\det P=1$ which is
  usually parametrized as
  \begin{align}
    \label{phase}
    P = e^{i\,\varphi}\,\diag (e^{i\phi_{13}},e^{i\phi_{23}},1)\ .
  \end{align}
  These phases cannot be absorbed by field redefinitions
  \cite{ellis79}.
  The CKM matrix is then defined as
  \begin{align}
    V_\ckm = U_u^\dagger\,U_d\ .
  \end{align}
  The most general weak basis transformation which leaves the
  interactions invariant is:
  \begin{align*}
    \mathsf{10}_i 
    & \to \mathsf{10}^{\prime}_j = \Ucal_{ij}\mathsf{10}_j\ , \\ 
    \mathsf{5}^*_i 
    & \to \mathsf{5}^{*\,\prime}_j = \Vcal_{ij}\mathsf{5}^*_j\ .
  \end{align*} 
  Then the Yukawa matrices transform like 
  \begin{align*}
    Y_u & \to \Ucal^\top\, Y_u\, \Ucal\ ,\\
    Y_d & \to \Ucal^\top\, Y_d\, \Vcal\ .
  \end{align*}
  
  The superpotential of the {\sf SU(5)} Yukawa interactions expressed
  in terms of SM superfields is given by
  eqn.~(\ref{SM-superpotential}).  Transforming the singlets fields by
  $\Phi\to\Wcal_\Phi \Phi$, the superpotential transforms like
  \begin{align*}
      W_Y = \;
      & Q^\top\, (\Ucal^\top\,Y_u\,\Ucal\,\Wcal_u)\, u^\cc\,
      H_f 
      + Q^\top\, (\Ucal^\top\,Y_d\,\Vcal\,\Wcal_d)\, d^\cc\,
      \overline{H}_f
      + e^{\cc\top}\,
      (\Wcal_e^\top\,\Ucal^\top\,Y_e\,\Vcal)\, L\, \overline{H}_f\\  
      & + \tfrac{1}{2}\, Q^\top\, (\Ucal^\top\,Y_{qq}\,\Ucal)\, Q\,
      H_\cc 
      + Q^\top\, (\Ucal^\top\,Y_{ql}\,\Vcal)\, L\,
      \overline{H}_\cc \\
      & + u^{\cc\top}\,
      (\Wcal_u^\top\,\Ucal^\top\,Y_{ue}\,\Ucal\,\Wcal_e)\, e^\text{\sc
        c}\, H_\cc
      + u^{\cc\top}\,
      (\Wcal_u^\top\,\Ucal^\top\,Y_{ud}\,\Vcal\,\Wcal_d)\, d^\text{\sc
        c}\, \overline{H}_\cc
      \ .
  \end{align*}


  There are two possible physical bases now, 
  namely diagonal up quark and diagonal down quark matrices,
  which can be realized by a suitable choice of all transformation
  matrices. 
  With $Y_{qq}=Y_{ue}=Y_u$ and $Y_{ql}=Y_{ud}=Y_d$,
  the Yukawa interactions 
  read 
  \begin{align}
    \begin{split}
      W_Y = \;
      & Q^\top\, \Dcal_u\, u^\cc\, H_f 
      + Q^\top\, (V_\ckm\, \Dcal_d)\, d^\cc\,
      \overline{H}_f  
      + e^{\cc\top}\, \Dcal_e\, L\, \overline{H}_f\\ 
      & + \tfrac{1}{2}\, Q^\top\, (\Dcal_u\,P)\, Q\, H_\cc 
      + Q^\top\, (V_\ckm\,\Dcal_d)\, L\, \overline{H}_\text{\sc
        c} \\
      & + u^{\cc\top}\, (\Dcal_u\,V_\ckm^*)\,
      e^\cc\, H_\cc
      + u^{\cc\top}\, (P^*\,V_\ckm\,\Dcal_d)\,
      d^\cc\, \overline{H}_\cc
    \end{split}
    \label{hisano-basis}
  \end{align}
  in the first and
  \begin{align}
    \begin{split}
      W_Y = \;
      & Q^\top\, (V_\ckm^\dagger\,\Dcal_u)\, u^\cc\,
      H_f 
      + Q^\top\, \Dcal_d\, d^\cc\, \overline{H}_f 
      + e^{\cc\top}\, \Dcal_e\, L\, \overline{H}_f\\
      & + \tfrac{1}{2}\, Q^\top\, (V_\text{\sc
        ckm}^\dagger\,\Dcal_u\,P\,V_\ckm^*)\, Q\, H_\cc
      + Q^\top\, \Dcal_d\, L\, \overline{H}_\cc\\
      & + u^{\cc\top}\, (\Dcal_u\,V_\ckm^*)\,
      e^\cc\, H_\cc 
      + u^{\cc\top}\, (P^*\,V_\ckm\,\Dcal_d)\,
      d^\cc\, \overline{H}_\cc
    \end{split}
    \label{goto-basis}
  \end{align}
  in the second basis.
  The former is used in ref.~\cite{hisano93},
  the latter in ref.~\cite{goto99}.

  In principle, these formulae are only valid for unbroken
  supersymmetry where one can use the same transformations for the
  fermions and their supersymmetric partners.
  Broken supersymmetry gives small corrections to these
  transformations \cite{nath85}.

  \subsubsection*{Consistent model}

  Expanding the superpotential by higher dimensional operators,
  the identities (\ref{up-unification}) and (\ref{down-unification}), 
  \begin{gather*}
    Y_u = Y_{qq} = Y_{ue} = Y_1 \ , \qquad
    Y_d = Y_{ql} = Y_{ud} = Y_2 \ ,
  \end{gather*}
  at $M_\text{GUT}$ no longer hold.
  Instead, 
  one can derive the following relations between the matrices:
  \begin{align}
    \begin{split}
      Y_{qq} - Y_{ue} & =
      \frac{5}{2}\frac{\sigma}{M_\text{Pl}}f_2^A \ , \\
      Y_u - Y_{qq} & =
      5\frac{\sigma}{M_\text{Pl}}f_1^S  
      + \frac{5}{4}\frac{\sigma}{M_\text{Pl}}\left(f_2^S+f_2^A\right)
      \ , \\ 
      Y_u - Y_{ue} & =
      5\frac{\sigma}{M_\text{Pl}}f_1^S  
      + \frac{5}{4}\frac{\sigma}{M_\text{Pl}}\left(f_2^S+3f_2^A\right)
      \ , \\
      Y_d - Y_e = Y_{ud}-Y_{ql}
      & = 5\frac{\sigma}{M_\text{Pl}}h_2 
      \ , \\ 
      \tfrac{3}{5} Y_d + \tfrac{2}{5} Y_e  
      & = Y_2 -3\frac{\sigma}{M_\text{Pl}}h_1 \ , \\
      Y_{ql} - Y_e = Y_{ud} - Y_d  
      & = 5\frac{\sigma}{M_\text{Pl}} h_1
      \ . 
    \end{split}
    \label{consistent-relations}
  \end{align}
  The antisymmetric part of $f_2$ is determined by the difference
  between $Y_{qq}$ and $Y_{ue}$,
  then only symmetric terms of $f_1$ and $f_2$ remain.


  \section{Renormalization group equations
    \label{rge}}

  \subsubsection*{Yukawa couplings}
  
  The one-loop renormalization group equations, in the
  $\overline{\text{\rm MS}}$ scheme, can be written for general 
  Yukawa matrices \cite{rge}
  \begin{align}
    \begin{split}
      16\pi^2 \frac{d\, Y_u}{dt}&=
      \left[\,T_u-G_u(t)+\frac32\left(b\,Y_u\,Y_u^{\dagger}+
        c\,Y_d\,Y_d^{\dagger}\right)\right]Y_u\
      ,\\ 16\pi^2 \frac{d\, Y_d}{dt}&=
      \left[\,T_d-G_d(t)+\frac32\left(b\,Y_d\,Y_d^{\dagger}+
        c\,Y_u\,Y_u^{\dagger}\right)\right]Y_d\
      ,\\ 16\pi^2 \frac{d\, Y_e}{dt}&=
      Y_e\,\left(\,T_e-G_e(t)+\frac32b\,Y_e^{\dagger}\,Y_e\right)\ ,
    \end{split}
    \label{rge-formulae}
  \end{align}
  where $t=\log \mu/M_Z$ and the traces $T_u, T_d, T_e$ are given by 
  \begin{align}
    \begin{split}
      T_u&=\trace\,(3\,Y_u\,Y_u^{\dagger}+3\,a\,Y_d\,Y_d^{\dagger}+
      a\,Y_e^{\dagger}\,Y_e)\ ,\\
      T_d=T_e&=\trace\,(3\,a\,Y_u\,Y_u^{\dagger}+3\,Y_d\,Y_d^{\dagger}+
      \,Y_e^{\dagger}\,Y_e)\ .
    \end{split}
    \label{rge-formulae2}
  \end{align}
  The constants $a$, $b$ and $c$ as well as the functions $G_u(t)$,
  $G_d(t)$ and $G_e(t)$, are summarized in the 
  table\,\ref{tab:def}. 
  
  \begin{table}[h]
    \centering
    \begin{tabular}{lll}
      & SM & MSSM \\
      \hline
      \phantom{$\cfrac{1}{2}$}
      (a, b, c) $\quad$ & (1, 1, $-\frac{3}{2}$) & (0, 2, 1) \\
      \hline
      \phantom{$\cfrac{1}{2}$} $G_u(t)$ & 
      $\tfrac{17}{20}\,g_1^2(t)+\tfrac{9}{4}\,g_2^2(t)+8\,g_3^2(t)
      \qquad $ & 
      $\tfrac{13}{15}\,g_1^2(t)+3\,g_2^2(t)+\tfrac{16}{3}\,g_3^2(t)$
      \\
      \phantom{$\cfrac{1}{2}$} $G_d(t)$ &
      $\tfrac{1}{4}\,g_1^2(t)+\tfrac{9}{4}\,g_2^2(t)+8\,g_3^2(t)$ &
      $\tfrac{7}{15}\,g_1^2(t)+3\,g_2^2(t)+\tfrac{16}{3}\,g_3^2(t)$ \\
      \phantom{$\cfrac{1}{2}$} $G_e(t)$ &
      $\tfrac{9}{4}\,g_1^2(t)+\tfrac{9}{4}\,g_2^2(t)$ &
      $\tfrac{9}{5}\,g_1^2(t)+3\,g_2^2(t)$ \\
      \hline
      \phantom{$\cfrac{1}{2}$} $b_1$ &
      $\tfrac43n+\tfrac{1}{10}m$ & $2\,n+\tfrac3{10}m$ \\
      \phantom{$\cfrac{1}{2}$} $b_2$ &
      $\tfrac43n+\tfrac{1}{6}m-\tfrac{22}3$ & $2\,n+\tfrac12m-6$ \\ 
      \phantom{$\cfrac{1}{2}$} $b_3$ &
      $\tfrac43n-11$ & $2\,n-9$ \\
      \hline
    \end{tabular}
    \caption{Coefficients to (\ref{rge-formulae}) and
      (\ref{rge-formulae2}).
      The running gauge coupling constant at 1-loop is given by
      $g^2_i(t)=g^2_i(0)/\left(1-\frac{b_i}{8\pi^2}\,g^2_i(0)\,t\right)$.
      The integers $n$ and $m$ stand for number of generations and
      Higgs doublets, respectively.}
    \label{tab:def}
  \end{table}
  
  The equations for the Wilson coefficients read \cite{goto99}
  \begin{align}
    \begin{split}
      \label{c5Lren}
      16\pi^2 \frac{d}{dt} C_{5L}^{ijkl} 
      & = \left( 
        -8\,g_3^2 -6\,g_2^2 -\tfrac{2}{3}g_1^2 \right) C_{5L}^{ijkl} 
      + C_{5L}^{mjkl} \! \left( \!
        Y_d Y_d^\dagger + Y_u Y_u^\dagger \right)^{i}_{m} \\
      & \quad + C_{5L}^{imkl} \! \left( 
        Y_e^\dagger Y_e \right)_{m}^{j} \! \!
      + C_{5L}^{ijml} \! \left( \! 
        Y_d Y_d^\dagger + Y_u Y_u^\dagger \right)^{k}_{m} \! \!
      + C_{5L}^{ijkm} \! \left( \! 
        Y_d Y_d^\dagger + Y_u Y_u^\dagger \right)^{l}_{m} ,
    \end{split}  \\
    \begin{split}
      \label{c5Rren}
      16\pi^2 \frac{d}{dt} C_{5R}^{ijkl} 
      & =  \left(
        -8\,g_3^2 -4\,g_1^2 \right) C_{5R}^{ijkl}
      + C_{5R}^{mjkl} \left( 
        2\, Y_u^\dagger Y_u \right)_{m}^{i} \\
      & \quad + C_{5R}^{imkl} \left( 
        2\, Y_d^\dagger Y_d \right)_{m}^{j}
      + C_{5R}^{ijml} \left( 2\, Y_e Y_e^\dagger \right)^{k}_{m} 
      + C_{5R}^{ijkm} \left( 2\, Y_u^\dagger Y_u \right)_{m}^{l} .
    \end{split}
  \end{align}

  \subsubsection*{Gauge couplings, Constraint on $M_{H_\cc}$}

  In minimal supersymmetric {\sf SU(5)},
  the RGE at one-loop level are given by \cite{hisano93}:
  {\footnotesize
    \begin{align*} 
      \alpha_{1}^{-1}(M_Z)= 
      \alpha_5^{-1}(\Lambda) + \frac{1}{2\pi} \Blb
      \left( -\frac{2}{3}n-\frac{1}{2} \right) 
      \log \frac{M_S}{M_Z}
      + \left( 2n+\frac{3}{5} \right) \log\frac{\Lambda}{M_Z} 
      -10 \log \frac{\Lambda}{M_V} 
      + \frac{2}{5} \log \frac{\Lambda}{M_{H_\cc}} 
      \Brb ,
    \end{align*}
    \begin{align*}
      \alpha_{2}^{-1}(M_Z)=
      \alpha_5^{-1}(\Lambda) + \frac{1}{2 \pi} \Blb
      \left( -\frac{2}{3}n -\frac{13}{6} \right) 
      \log \frac{M_S}{M_Z} 
      + \left( 2n-5 \right) \log \frac{\Lambda}{M_Z} 
      -6 \log \frac{\Lambda}{M_V} 
      + 2\log \frac{\Lambda}{M_3} 
      \Brb ,
    \end{align*}
    \begin{align*}
      \alpha_{3}^{-1}(M_Z)\!=\!
      \alpha_5^{-1}(\Lambda) + \frac{1}{2 \pi} \Blb
      \left( -\frac{2}{3}n\!-\!2 \right) 
      \log \frac{M_S}{M_Z}
      + \left( 2n\!-\!9 \right)\log \frac{\Lambda}{M_Z}
      -4 \log \frac{\Lambda}{M_V}  
      +3 \log \frac{\Lambda}{M_8} 
      + \log \frac{\Lambda}{M_{H_\cc}} 
      \Brb ,
    \end{align*}
    }
  \noindent
  where $M_S$ is the SUSY breaking scale.
  Using the combinations 
  \begin{align}
    (-\alpha_1^{-1}+3\alpha_2^{-1}-2\alpha_3^{-1})\; (M_Z)
    &= \frac{1}{2\pi} \left[
      - 2 \, \log \frac{M_S}{M_Z}  
      + \frac{12}{5} \log \left(
        \frac{M_{H_\cc}}{M_Z} \left(
          \frac{M_3}{M_8}
        \right)^\frac{5}{2}
      \right)
    \right] \ , 
    \label{MHC} \\
    (5\alpha_1^{-1}-3\alpha_2^{-1}-2\alpha_3^{-1})\; (M_Z) 
    &= \frac{1}{2\pi} \left[
      8 \, \log \frac{M_S}{M_Z}
      + 12 \, \log \frac{
        M_V^2 \sqrt{M_3 M_8}
        }{M_Z^3}
    \right] 
    \label{MGUT}
  \end{align}
  one can derive constraints on the products 
  $M_{H_\cc} (M_3/M_8)^{5/2}$ and 
  $M_V^2 \sqrt{M_3 M_8} \equiv M_\text{GUT}^3$. 
  At two-loop level,
  there are no simple analytic relations any more.

  Taking $M_8 = M_3 = M_\Sigma$
  as it was done for the constraint in eqs.~(\ref{newconstraint})
  and (\ref{gutconstraint}),
  it simply reads $M_{H_\cc}$ and
  $M_\text{GUT}^3=M_V^2 M_\Sigma$.



  \section{Diagrams 
    \label{diagrams}}

  \begin{figure}
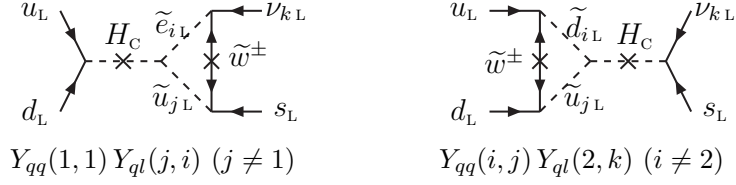
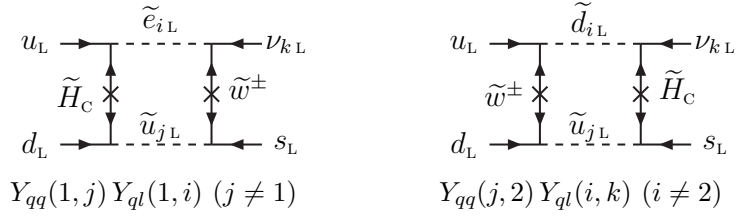
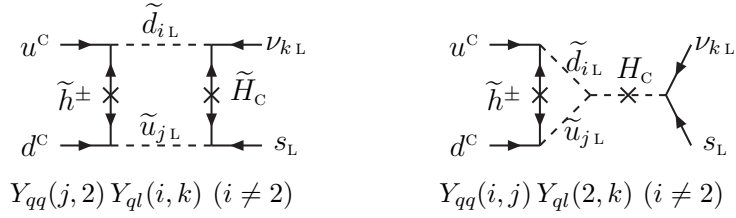
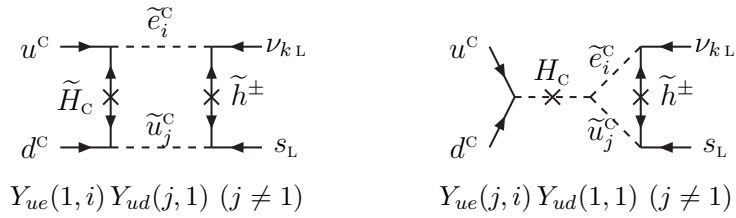

    \centering
    
    \subfigure[$LLLL$ triangle diagrams]
    { 
      \label{chargino-exchange:a}
      \gdef\boson{w}
      \finalstate{u_\le}{d_\le}{s_\le} 
      \triud{e_{i\,\le}}{u_{j\,\le}}{Y_{qq}(1,1)\,Y_{ql}(j,i)\
        (j\ne1)}
      \finalstate{u_\le}{d_\le}{s_\le}
      \cistriud{d_{i\,\le}}{u_{j\,\le}}{Y_{qq}(i,j)\,Y_{ql}(2,k)\
        (i\ne2)}
      }
    \subfigure[$LLLL$ box diagrams]
    { 
      \label{chargino-exchange:b}
      \gdef\boson{w}
      \finalstate{u_\le}{d_\le}{s_\le} 
      \boxud{e_{i\,\le}}{u_{j\,\le}}{Y_{qq}(1,j)\,Y_{ql}(1,i)\
        (j\ne1)}
      \finalstate{u_\le}{d_\le}{s_\le} 
      \cisboxud{d_{i\,\le}}{u_{j\,\le}}{Y_{qq}(j,2)\,Y_{ql}(i,k)\
        (i\ne2)}
      }
    \subfigure[$RRLL$ diagrams]
    {
      \label{chargino-exchange:c}
      \gdef\boson{h}
      \finalstate{u^\cc}{d^\cc}{s_\le} 
      \cisboxud{d_{i\,\le}}{u_{j\,\le}}{Y_{qq}(j,2)\,Y_{ql}(i,k)\
        (i\ne2)}
      \finalstate{u^\cc}{d^\cc}{s_\le}
      \cistriud{d_{i\,\le}}{u_{j\,\le}}{Y_{qq}(i,j)\,Y_{ql}(2,k)\
        (i\ne2)}
      }
    \subfigure[$RRRR$ diagrams]
    {
      \label{chargino-exchange:d}
      \gdef\boson{h}
      \finalstate{u^\cc}{d^\cc}{s_\le}
      \boxud{e_i^\cc}{u_j^\cc}{Y_{ue}(1,i)\,Y_{ud}(j,1)\ (j\ne1)}
      \finalstate{u^\cc}{d^\cc}{s_\le}
      \triud{e_i^\cc}{u_j^\cc}{Y_{ue}(j,i)\,Y_{ud}(1,1)\ (j\ne1)}
      }
    \caption{Diagrams with chargino dressing}
    \label{chargino-exchange}
  \end{figure}

  \gdef\charge{0}

  \begin{figure}
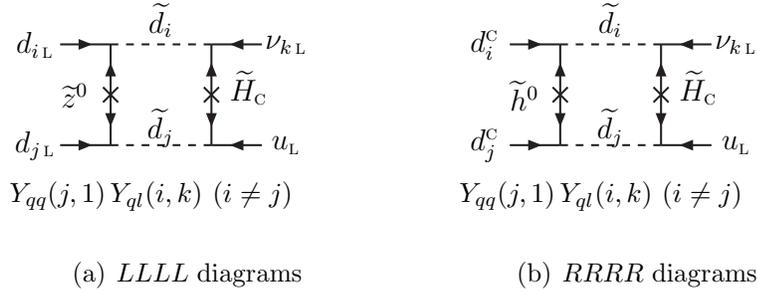

    \centering

    \hspace{-15mm}

    \subfigure[$LLLL$ diagrams]{
      \label{neutralino-exchange:a}
      \gdef\boson{z}
      \finalstate{d_{i\,\le}}{d_{j\,\le}}{u_\le}
      \cisboxud{d_i}{d_j}{Y_{qq}(j,1)\,Y_{ql}(i,k)\ (i\ne j)}}
    \subfigure[$RRRR$ diagrams]{
      \label{neutralino-exchange:b}
      \gdef\boson{h}
      \finalstate{d^\cc_i}{d^\cc_j}{u_\le}
      \cisboxud{d_i}{d_j}{Y_{qq}(j,1)\,Y_{ql}(i,k)\ (i\ne j)}}
    \caption{Diagrams with neutralino dressing}
    \label{neutralino-exchange}
  \end{figure}

  Fig. \ref{chargino-exchange} lists the diagrams for the decay 
  $p\to K^+ \bar\nu$ with chargino dressing;
  they are related to the first addend in (\ref{amplitude}). 
  Those with right-handed fermions incoming
  can be divided into two groups depending on the dressing before 
  (fig.~\ref{chargino-exchange:c})
  or after the decay operator (fig.~\ref{chargino-exchange:d}); 
  we therefore call them $RRLL$ and $RRRR$ diagrams, respectively.
  The latter case is the only one related to the $RRRR$ operator 
  and $C_{5R}$
  because there are no right-handed neutrinos in the model.
  %
  As discussed in Section \ref{dim5},
  the dimension five operators are flavour non-diagonal,
  hence several diagrams are suppressed.

  By interchanging down and strange quarks as incoming and outgoing
  particles, we get the diagrams due to the second addend.
  The diagrams of the last one cannot be realized by chargino
  exchange,
  so we look at those with neutralino exchange that are given in
  fig.~\ref{neutralino-exchange}.

  The coefficients $C_{LL}$ and $C_{RL}$ used in
  eqn.~(\ref{amplitude}) then read 
  \begin{equation}
    \label{coefficient}
    C_{LL/RL} = \frac{\Kcal}{M_{H_\cc}}\, 
      C_{5L/5R}\, A_s \, A_l\, f(M;m_1,m_2) 
  \end{equation}
  where $C_{5L}=Y_{qq} Y_{ql}$ is related to $C_{LL}$ and 
  $C_{5R}=Y_{ue} Y_{ud}$ to $C_{RL}$.
  They are evaluated at the GUT scale 
  and $A_s$ gives the correction due to running from GUT to
  SUSY breaking scale where $\Kcal$ is determined.
  In practice,
  the Wilson coefficients $C_5$ are renormalized by means of 
  eqn.~(\ref{c5Lren}) and (\ref{c5Rren}) and evaluated at SUSY
  breaking scale. 
  Finally,
  $A_l$ describes the renormalization of the $d=6$ coefficients down
  to 1\,GeV.


  \section{Renormalization factors 
    \label{factor}} 

  The renormalization factors are crucial for analyzing the proton
  decay.
  Since there is some discrepancy in the literature,
  we want to discuss them here in detail.
  
  As already mentioned in Section \ref{dim5}, there are two ranges for
  the renormalization, namely the short-distance between
  $M_\text{GUT}$ and $M_\text{SUSY}$ and the long-distance between the
  latter and the proton mass at $\sim$ 1\,GeV leading to the factors
  $A_s$ and $A_l$.  The former is highly dependent on the top Yukawa
  coupling $y_t$ and can therefore not be calculated analytically.
  
  The renormalization group effects in SUSY GUTs have first been
  discussed in ref.~\cite{ellis82}.  At that time, not only the high
  top mass was unknown ($m_t=20\,$GeV was assumed), but since there
  were no data at $M_Z$, the values at 1\,GeV were taken to calculate
  the decay rate.  Hence the renormalization factors $A_S$ and $A_L$
  were defined, which include the running factor of the Yukawa
  couplings from low to high scale.  In this work, we use the Yukawa
  couplings at $M_Z$ and $M_\text{SUSY}$ and evaluate their values at
  $M_\text{GUT}$.  These are taken as input parameters for the
  calculation, so our factors $A_s$ and $A_l$ differ from $A_S$ and
  $A_L$ in ref.~\cite{ellis82,hisano93}.  For the long-distance part,
  this discrepancy was stressed in ref.~\cite{dermisek}.
  
  Because of the high top mass, $A_s$ cannot be solved analytically
  \cite{hisano93} and depends on the related particles.  Hence the
  Wilson coefficients are evolved down to $M_\text{SUSY}$ by using
  eqs.~(\ref{c5Lren}) and (\ref{c5Rren}).  For simplicity,
  $M_\text{SUSY}$ is identified with the electroweak scale, so $A_l$
  describes pure QCD renormalization down to 1\,GeV\footnote{The
    authors of ref.~\cite{dermisek} obtain $A_l=1.32$.},
  \begin{align}
    A_l =
    \Blb \frac{\alpha_3(\mu_\text{had})}{
      \alpha_3(M_Z)}\Brb^\frac{6}{33-2n_{\!f}}
    \to \Blb \frac{\alpha_3(\mu_\text{had})}{
      \alpha_3(m_c)}\Brb^\frac{2}{9}
    \Blb \frac{\alpha_3(\mu_\text{c})}{
      \alpha_3(m_b)}\Brb^\frac{6}{25}
    \Blb \frac{\alpha_3(m_b)}{
      \alpha_3(m_Z)}\Brb^\frac{6}{23}
    = 1.43 \ ,
    \label{al}
  \end{align}
  with $\mu_\text{had}=1$\,GeV.


\end{appendix}

\end{document}